# Interstellar Neutral Hydrogen in the Heliosphere:
# New Horizons Observations in the Context of Models

P. Swaczyna[1,*], M. Bzowski[1], K. Dialynas[2], L. Dyke[3], F. Fraternale[4], A. Galli[5], J. Heerikhuisen[6], M. Z. Kornbleuth[7], D. Koutroumpa[8], I. Kowalska-Leszczyńska[1], M. A. Kubiak[1], A. T. Michael[9], H.-R. Müller[3], M. Opher[7], F. Rahmanifard[10]

[1]Space Research Centre PAS (CBK PAN), Bartycka 18a, 00-716 Warsaw, Poland
[2]Center for Space Research & Technology, Academy of Athens, 10679 Athens, Greece
[3]Department of Physics and Astronomy, Dartmouth College, Hanover, NH 03755, USA
[4]Center for Space Plasma and Aeronomic Research, The University of Alabama in Huntsville, Huntsville, AL 35899, USA
[5]Space Research and Planetary Sciences, Physics Institute, University of Bern, 3012 Bern, Switzerland
[6]Department of Mathematics and Statistics, University of Waikato, Hamilton 3240, New Zealand
[7]Astronomy Department, Boston University, Boston, MA 02215, USA
[8]LATMOS-IPSL, CNRS, UVSQ Paris-Saclay, Sorbonne Université, 78280 Guyancourt, France
[9]Johns Hopkins University Applied Physics Laboratory, Laurel, MD 20723, USA
[10]Physics Department, Space Science Center, University of New Hampshire, Durham, NH 03824, USA

**Abstract**

Interstellar neutral (ISN) hydrogen is the most abundant species in the outer heliosheath and the very local interstellar medium (VLISM). Charge exchange collisions in the outer heliosheath result in filtration, reducing the ISN hydrogen density inside the heliosphere. Additionally, these atoms are intensively ionized close to the Sun, resulting in a substantial reduction of their density within a few au from the Sun. The products of this ionization – pickup ions (PUIs) – are detected by charged particle detectors. The Solar Wind Around Pluto (SWAP) instrument on New Horizons provides, for the first time, PUI observations from the distant heliosphere. We analyze the observations collected between 22 and 52 au from the Sun to find the ISN hydrogen density profile and compare the results with predictions from global heliosphere models. We conclude that the density profile derived from the observations is inconsistent with steady-state model predictions. This discrepancy is not explained by time variations close to the Sun and thus may be related to the temporal evolution of the outer boundaries or VLISM conditions. Furthermore, we show that the cold and hot models of ISN hydrogen distribution are not a good approximation closer to the termination shock. Therefore, we recommend a new fiduciary point based on the available New Horizons observations at 40 au from the Sun, at ecliptic direction (285.62°, 1.94°), where the ISN hydrogen density is 0.11 cm$^{-3}$. The continued operation of New Horizons should give better insight into the source of the discussed discrepancy.

## 1. Introduction

Hydrogen atoms are the most abundant species in the very local interstellar medium (VLISM) around the heliosphere (Frisch et al. 2011). Because the Sun moves relative to the VLISM, interstellar neutral (ISN) atoms flow into the heliosphere (Axford et al. 1963). The ISN atoms are coupled to the magnetized plasma through charge exchange collisions (Ruciński et al. 1996). The mean free path between charge exchange collisions in the VLISM is of the order of ~100 au, i.e., comparable with the size of the heliosphere. Consequently, while the collisions are not frequent enough to equilibrate the conditions between the plasma and neutrals, they impact the density distribution of ISN hydrogen around and in the heliosphere. Moreover, the ISN hydrogen population in the heliosphere is slower and warmer than the pristine VLISM conditions (Ripken & Fahr 1983; Lallement & Bertaux 1990).

The plasma density is significantly lower in the heliosphere than in the VLISM and outer heliosheath, making charge exchange less frequent (Gurnett & Kurth 2019; Richardson et al. 2019). The ionized ISN atoms have a distinctive energy in the plasma frame and are called pickup ions (PUIs, Möbius et al. 1985). Furthermore, neutralized energetic ions, such as pickup ions, form energetic neutral atoms (ENAs). Even though ENAs contribute to the neutral hydrogen density in the heliosphere, their density is orders of magnitude lower than that of the ISN hydrogen (Bzowski & Galli 2019). PUIs are also created through photoionization and electron impact ionization (Bzowski et al. 2013). The region inside which the ISN density is reduced by these ionization processes more than $1/e$ is called the ionization cavity. The cavity evolves over the solar cycle, extending to ~4-5 au from the Sun in the upwind direction and more than 20 au on the downwind side (Ruciński & Bzowski 1995; Sokół et al. 2019b).

---

[*] Corresponding author (pswaczyna@cbk.waw.pl)

The filtration in the outer heliosphere and filtration close to the Sun require different modeling approaches. The first one is typically addressed by global modeling of the heliosphere using a kinetic model of the neutral component, self-consistently coupled with a magnetohydrodynamic (MHD) description for plasma (e.g., Malama 1991; Baranov & Malama 1993; Heerikhuisen et al. 2006). The global approach to modeling of the ISN hydrogen typically does not provide good statistics close to the Sun. Moreover, many of these models do not consider details of the solar cycle evolution of ionization and radiation pressure due to computational limitations. To overcome this problem, the modeling in the proximity of the Sun uses time- and heliolatitude-dependent models evolved from the cold and hot models of the ISN gas in the heliosphere (Fahr 1968; Thomas 1978; Ruciński & Bzowski 1995; Tarnopolski & Bzowski 2009) based on boundary conditions at the termination shock.

The observations performed close to the Sun, i.e., within the ionization cavity, require detailed modeling of the ionization processes, including the temporal and latitudinal variations (Sokół et al. 2019a), and thus are prone to significant uncertainties. Therefore, observations from larger distances are crucial to derive the ISN hydrogen properties. For example, the density can be derived from observations of PUIs in the outer heliosphere (Zirnstein et al. 2022). The PUI observations from Ulysses from distances of ~5 au allowed Bzowski et al. (2008) to derive the ISN hydrogen density at the termination shock density of ~0.09 cm$^{-3}$. This value was consistent with estimations using other methods (Pryor et al. 2008; Richardson et al. 2008; Bzowski et al. 2009). However, Dialynas et al. (2019) utilized a unique combination of 28-3,500 keV *in situ* ions from Voyager 2 and 5.2-55 keV ENAs from INCA on Cassini and predicted that the ISN hydrogen density should be ~30% higher (~0.12 cm$^{-3}$) beyond the termination shock. Moreover, Swaczyna et al. (2020) found that the New Horizons observations performed up to ~38 au from the Sun, i.e., far beyond the ionization cavity, indicate the density higher by ~40%. This paper provides an update on the derivation of the ISN hydrogen density using New Horizons PUI observations up to ~52 au from the Sun (McComas et al. 2021, 2022) and a comparison of the results with the predictions of the global heliosphere models.

## 2. PUI Observations from New Horizons

NASA's New Horizons mission was primarily intended to study the Pluto system and the Kuiper belt (Stern 2008). The spacecraft was placed on a trajectory escaping the solar system using Jupiter's gravity assist. Solar Wind Around Pluto (SWAP), a plasma instrument of the mission, was designed to measure the rarefied solar wind plasma at large distances from the Sun (McComas et al. 2008). SWAP collects ions entering the detector from a wide (276°×10°) field of view in narrow ($\Delta E/E$ = 0.085 FWHM) energy per charge bins from 35 eV/q to 7.5 keV/q. This energy range covers solar wind protons, α-particles, and PUIs created in the supersonic solar wind from ISN hydrogen atoms. In the early stage of the mission, the SWAP was on only during short intervals (McComas et al. 2010; Randol et al. 2012, 2013). Since 2012, SWAP operates most of the time, providing almost continuous observations starting at ~22 au from the Sun. The top panel of Figure 1 shows the New Horizons trajectory projected on the ecliptic plane. The spacecraft remains close to the ecliptic plane. Between 22 and 52 au (the range of radial distances used in our analysis), the spacecraft's ecliptic longitude changed slowly from 280.03° to 287.32°, while the ecliptic latitude changed from 1.81° to 1.98°.

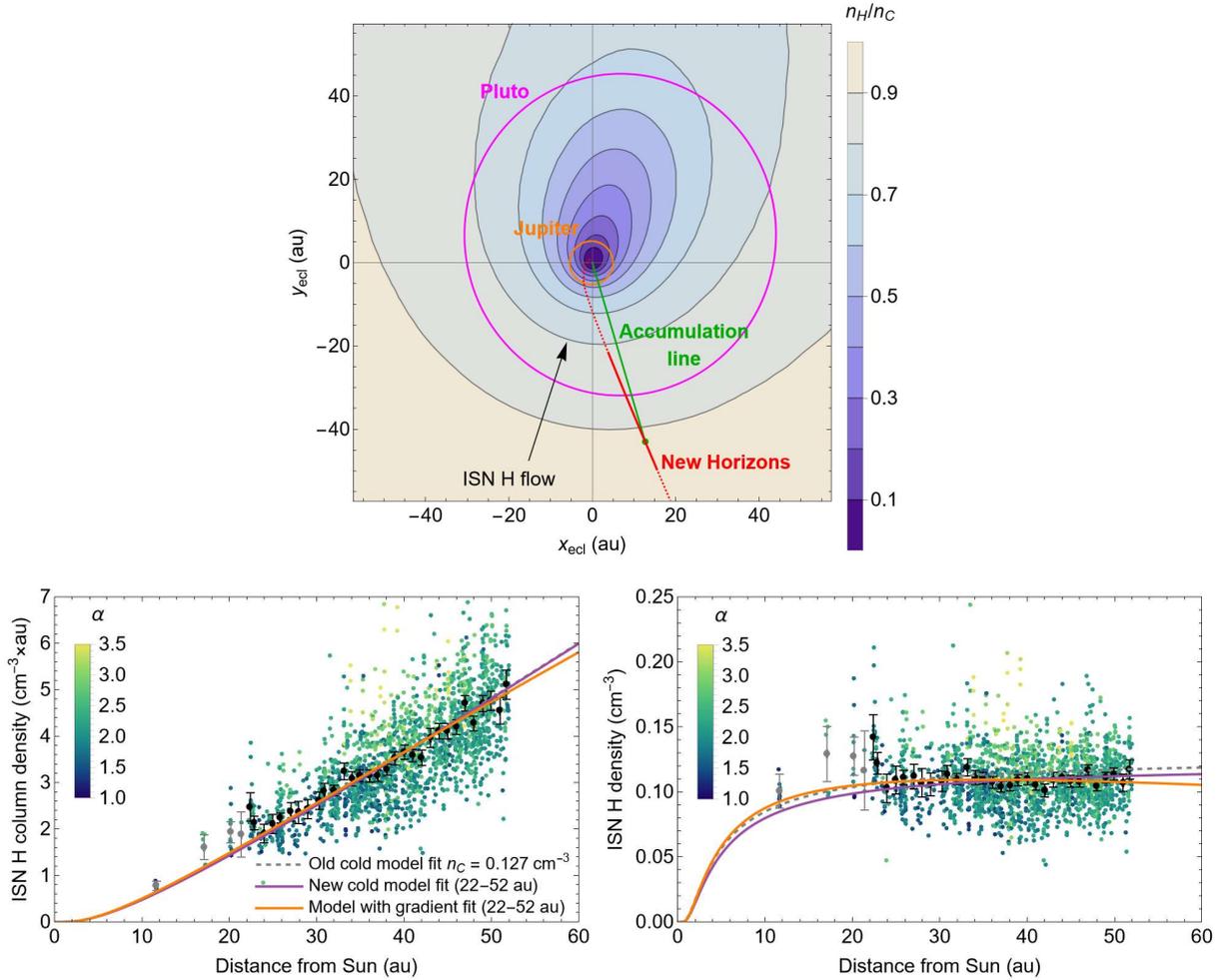

**Figure 1.** *Top panel*: the trajectory of New Horizons (red line) projected on the ecliptic plane shown on top of the density modulation of ISN hydrogen calculated with the Warsaw Test Particle Model (WTPM, Tarnopolski & Bzowski 2009; Sokół et al. 2015). The solid portion represents the almost continuous observations between 22 and 52 au used in our study. The orange and magenta ellipses show the Jupiter and Pluto trajectories, respectively. The black arrow indicates the ISN hydrogen flow direction. The green line represents the line along which the PUIs accumulated for observations at 45 au from the Sun. *Bottom panels*: column density (*left panel*) and local density (*right panel*) of ISN hydrogen in the heliosphere based on the New Horizons/SWAP observations of PUIs. The color of the data points corresponding to each 1 day of SWAP observations represents the cooling index value (see Swaczyna et al. 2020). The averages over 1-au-wide bins are shown using points with error bars (gray – excluded from the model fits, black – included in the model fits, see text). The cold model fit from Swaczyna et al. (2020) is shown with dashed gray lines. The new fit using a cold model and a model with a gradient (see text) are shown using purple and orange lines, respectively.

The SWAP PUI observations were analyzed in a series of papers. McComas et al. (2017) analyzed the PUI observations up to ~38 au by fitting the Vasyliunas & Siscoe (1976) distribution function. The methodology was later refined, accounting for solar wind $He^+$ ions near the PUI cutoff energy (Swaczyna et al. 2019a) and allowing for non-adiabatic cooling of PUIs according to the Chen et al. (2014) model (Swaczyna et al. 2020). McComas et al. (2021) found that fitting the PUI distribution cutoff velocity may further improve the goodness of the fit and analyzed the SWAP observations out to nearly 47 au. In February 2019, the SWAP flight software was updated to provide a higher cadence of PUI data, improving the time resolution from 1 day to ~30 minutes (McComas et al. 2022).

This paper updates the analysis by Swaczyna et al. (2020) with additional data (McComas et al. 2021, 2022). Because McComas et al. (2021) reprocessed all SWAP observations, the individual data points presented here may not overlap with those shown in Swaczyna et al. (2020). Nevertheless, the impact of these changes is negligible. Furthermore, we aggregate the data collected with the ~30-minute resolution into 1-day averages to match the earlier resolution. We select only those days with at least 42 half-hour data blocks. We apply the selection criteria discussed in Swaczyna et al. (2020).

The ISN hydrogen density profile may be derived from the PUI observations in the supersonic solar wind using two methods. The first method relates the PUI density with the column density of ISN hydrogen from the Sun to the distance at which the observations are collected. The ionization rate decreases with the distance $r$ from the Sun as $\beta(r) = \beta_0 (r_0/r)^2$, where $\beta_0$ is the ionization rate at $r_0 = 1$ au. The radially expanding solar wind advects the PUIs; thus, their accumulated flux in the solar wind is proportional to the column density:

$$u_\text{sw} n_\text{PUI}(r) = \left(\frac{r_0}{r}\right)^2 \beta_0 N_H(r), \qquad (1)$$

where $u_\text{sw}$ is the solar wind bulk speed, and $N_H(r) = \int_0^r n_H(r')\, dr'$ is the column density of the parent ISN hydrogen from the Sun to the distance $r$ (see green line in Figure 1, top panel). As discussed in detail in Swaczyna et al. (2020), we include corrections for the slowdown of the solar wind and the change in the proton density due to the PUI production through photoionization. The most statistically significant SWAP data points are two or three points immediately below the energy cutoff, dominated by freshly ionized PUIs. Therefore, the fit parameters may be used to derive the local density of ISN hydrogen. We closely follow the previous implementation, and the densities obtained from each day of the SWAP observations are presented in Figure 1. The individual data points present a significant scatter in the estimated densities caused by the solar wind variations. Therefore, we average them into 1-au-wide bins, each covering approximately 4 months of observations. The averaged densities are shown as black dots with error bars.

Swaczyna et al. (2020) found that the observed densities up to ~38 au are consistent with the cold model for the ISN hydrogen density far from the Sun of 0.127 cm$^{-3}$. The density of the ISN hydrogen in the cold model depends on the distance from the Sun $r$ and angular distance from the upwind direction $\theta$ (Thomas 1978):

$$n_H(r, \theta) = n_C \exp\left(-\frac{\lambda}{r}\frac{\theta}{\sin\theta}\right), \qquad (2)$$

where $n_C$ is the density of ISN hydrogen far from the Sun, and $\lambda$ is the size of the ionization cavity along the upwind direction. We adopt the direction of the ISN hydrogen inflow (252.2°, 9.0°) in ecliptic coordinates obtained from SWAN observations (Lallement et al. 2005). The adopted cavity size is consistent with the one used in the fitting of the PUI distribution function $\lambda = 4.0\pm0.5$ au (McComas et al. 2021; Sokół et al. 2019b). In fitting the models, we use only the observations collected from 22 au from the Sun, i.e., over the period of continuous SWAP observations. We removed the sporadic observations collected earlier because they represent shorter averaging times and thus may be biased by the solar wind variability. The best-fit densities far from the Sun $n_C$ are $0.1278\pm0.0043_{(\lambda)}\pm0.0011_{(\text{stat})}$ cm$^{-3}$ and $0.1217\pm0.0017_{(\lambda)}\pm0.0010_{(\text{stat})}$ cm$^{-3}$ for the method using the column density and local density, respectively. As in Swaczyna et al. (2020), we adopt the mean of the results and the discrepancy between these two results as a measure of methodological uncertainty. Therefore, the result is $n_C = 0.1248\pm0.0030_{(\text{method})}\pm0.0033_{(\lambda)}\pm0.0010_{(\text{stat})}$ cm$^{-3}$. This result is well within the uncertainty range reported by Swaczyna et al. (2020), which should be expected considering that the data set used here includes the data set from the previous paper. Swaczyna et al. (2020) discussed that the subtraction of He$^+$ ion contribution adds a 1.2% relative uncertainty ($\pm0.0015_{(\text{He+})}$ cm$^{-3}$), the adoption of the Lindsay & Stebbings (2005) charge exchange cross section adds a 10% relative uncertainty ($\pm0.013_{(\sigma)}$ cm$^{-3}$), and the instrumental uncertainty is ~4% ($\pm0.005_{(\text{instr})}$ cm$^{-3}$). Therefore, the combined uncertainty is $\pm0.015$ cm$^{-3}$.

Due to the systematic nature of the dominant uncertainties, our analysis does not reduce the combined uncertainty compared to the previous study. Nevertheless, the larger span in the distances from the Sun provides a means to study departures of the measured density profile from the density profile predicted by the cold model. In Section 3, we show that up to ~90 au from the Sun, the density profile can be described in all three considered models using the following *ad hoc* model with a gradient along the ISN hydrogen flow direction:

$$n_\text{H}(r, \theta) = n_\text{G}(1 + \gamma r \cos\theta) \exp\left(-\frac{\lambda}{r}\frac{\theta}{\sin\theta}\right), \qquad (3)$$

where $\gamma$ is a parameter describing the gradient slope along the inflow direction. Because the best-fit parameters obtained from separate fits to the column and local densities are strongly correlated, we perform a combined fit using the densities from both approaches. The best-fit parameters are $n_\text{G} = 0.138\pm0.008_{(\lambda)}\pm0.007_{(\text{stat})}$ cm$^{-3}$ and $\gamma = -0.0036\pm0.0010_{(\lambda)}\pm0.0011_{(\text{stat})}$ au$^{-1}$. Both these uncertainty components are almost perfectly anti-correlated, with a correlation coefficient of -0.98. Nevertheless, the fits to the cold model and the model with a gradient give $\chi^2 = 96.6$ and 64.2, respectively, using 62 data points. The other uncertainty sources mentioned above also impact parameter $n_\text{G}$ because they represent systematic effects that scale the densities. Combining all these sources of uncertainty yields a total uncertainty of $\pm 0.019$ cm$^{-3}$.

## 3. ISN Hydrogen Density in Global Heliosphere Models

In this paper, we use the ISN hydrogen densities obtained in three state-of-the-art hybrid kinetic-MHD global heliosphere models to put the New Horizons data in the context of global simulations. Model A has been developed over the years by the group at the University of Alabama in Huntsville (Pogorelov et al. 2008; Fraternale et al. 2021, 2023). Here, we use the most recent version, which incorporates hydrogen and helium atoms and ions treated self-consistently (see Fraternale et al. 2024). Model B is the product of the SHIELD Drive Science Center (Opher et al. 2015, 2017, 2020, 2023; Michael et al. 2021). Finally, Model C uses the same approach to the MHD solution as Model A but with different grid and boundary conditions (DeStefano & Heerikhuisen 2017; Heerikhuisen et al. 2019). Models A and C represent a traditional view of the heliosphere exhibiting a long tail in the downwind direction, while Model B challenges this view with a croissant-shaped heliosphere with the solar wind confined in two downwind jets. Table 1 summarizes the boundary conditions and selected assumptions employed in the models.

Figure 2 presents the ISN hydrogen density profiles along New Horizons' trajectory in the global heliosphere models. All profiles have the highest densities in the outer heliosheath corresponding to the so-called hydrogen wall (Baranov & Malama 1993). The models, which use steady-state conditions, are not optimized to reproduce the density of ISN hydrogen close to the Sun. Therefore, we do not expect them to reproduce the solution close to the Sun (see Section 4).

We fit the cold model formula using the model predictions between 22 and 52 au, i.e., for the same range as in the fit to New Horizons' data (Section 2), but we assume a fixed cavity size parameter $\lambda = 4$ au. The best-fit (asymptotic) ISN hydrogen densities far from the Sun $n_\text{C}$ are 0.119, 0.132, and 0.128 cm$^{-3}$ for models A, B, and C, respectively (purple lines in Figure 2). All models show a discrepancy in the radial gradient of the ISN hydrogen density. Therefore, we also fit the model given by Equation (3) to the global modeling results between 22 au and 90 au. The additional term is the simplest possible extension of the cold model. The factor $\cos\theta$ is motivated by the global distribution of ISN H, suggesting a large-scale gradient along the inflow direction (see, e.g., Heerikhuisen et al. 2008, Figure 1). The fitting procedure resulted in the following best-fit parameters: $n_\text{G} = 0.112$, 0.130, and 0.121 cm$^{-3}$, and $\gamma = 0.0018$, 0.0004, and 0.0018 au$^{-1}$ for models A, B, and C, respectively.

**Table 1. Boundary conditions and assumption in the global models**

|  | Model A | Model B | Model C |
|---|---|---|---|
| **Model description (reference)** | Fraternale et al. (2021, 2023, 2024) | Opher et al. (2015), Michael et al. (2021) | Heerikhuisen & Pogorelov (2010), Heerikhuisen et al. (2019) |
| **Simulation box** | (1680 au)$^3$, upwind boundary 680 au from Sun | (3000 au)$^3$, upwind boundary 1500 au from Sun | Sphere with a 1000 au radius around Sun |
| **Simulation grid resolution** | Adaptive$^a$ (10 au)$^3$ → (0.3 au)$^3$ | Adaptive$^{a,b}$ (6-12 au)$^3$ → (0.7 au)$^3$ | Adaptive$^a$ $r$: 10 au → 0.4 au Constant $\theta$: 1.5°, $\phi$: 3° |
| **LISM boundary conditions assumed at the outer boundary** | | | |
| **LISM flow speed** | 25.4 km s$^{-1}$ | 26.4 km s$^{-1}$ | 25.4 km s$^{-1}$ |
| **LISM flow direction (in ecliptic coord.)** | (75.7°, -5.1°) | (75.4°, -5.2°) | (75.7°, -5.1°) |
| **LISM temperature** | 7500 K | 6519 K | 7500 K |
| **LISM H$^0$ density** | 0.195 cm$^{-3}$ | 0.18 cm$^{-3}$ | 0.17 cm$^{-3}$ |
| **LISM He$^0$ density** | 0.0154 cm$^{-3}$ | … | 0.0153 cm$^{-3}$ |
| **LISM H$^+$ density** | 0.068 cm$^{-3}$ | 0.06 cm$^{-3}$ | 0.0494 cm$^{-3}$ |
| **LISM He$^+$ density** | 0.00898 cm$^{-3}$ | … | 0.0089 cm$^{-3}$ |
| **LISM magnetic field** | 2.93 µG | 4.4 µG | 2.93 µG |
| **LISM B-V angle** | 39.5° | 15.1° | 39.5° |
| **LISM B-V plane inclination to ecliptic** | 52.2° | 61.4° | 52.2° |
| **Solar wind (SW) boundary conditions** | | | |
| **Inner boundary at** | 10 au$^c$ | 30 au$^d$ | 10 au$^c$ |
| **SW speed** | 420 km s$^{-1}$ at 1 au | 417 km s$^{-1}$ at 30 au | 411 km s$^{-1}$ at 1 au |
| **SW density** | 8 cm$^{-3}$ at 1 au | 0.00874 cm$^{-3}$ at 30 au | 6.74 cm$^{-3}$ at 1 au |
| **SW composition** | H$^+$ + 3.5%$^e$ He$^{2+}$ | Only H$^+$ | H$^+$ + 4%$^e$ He$^{2+}$ |
| **SW temperature** | 80 000 K at 1 au | 108 680 K at 30 au$^f$ | 51 100 K at 1 au |
| **SW magnetic field** | Parker spiral $B_r$ = 37.5 µG at 1 au | Parker spiral $B_r$ = 65.5 µG at 1 au | Parker spiral $B_r$ = 37.5 µG at 1 au |
| **Model assumptions** | | | |
| **Plasma-neutral coupling** | Charge exchange + Photoionization | Charge exchange | Charge exchange + Photoionization |
| **He ions and atoms** | He$^{2+}$ in SW, He$^+$ in VLISM included self-consistently | … | He$^{2+}$ in SW, He$^+$ in VLISM comoving & thermalized with H$^+$ |
| **p-H$^0$ charge exchange** | <1 keV: Swaczyna et al. (2019b) >1 keV: Lindsay & Stebbings (2005) | Lindsay & Stebbings (2005) | <1 keV: Bzowski & Heerikhuisen (2020) >1 keV: Lindsay & Stebbings (2005) |

**Notes:** $^a$adaptive resolution with larger cells near the outer boundary and smaller cells near the inner boundary, $^b$grid for kinetic neutral component is coarser, see Michael et al. (2021) for details, $^c$analytic plasma solution within 10 au from the Sun is employed, no inner boundary for neutral atoms $^d$conditions advected from 1 au using the Izmodenov et al. (2005) model, $^e$by number, $^f$temperature of plasma with PUIs.

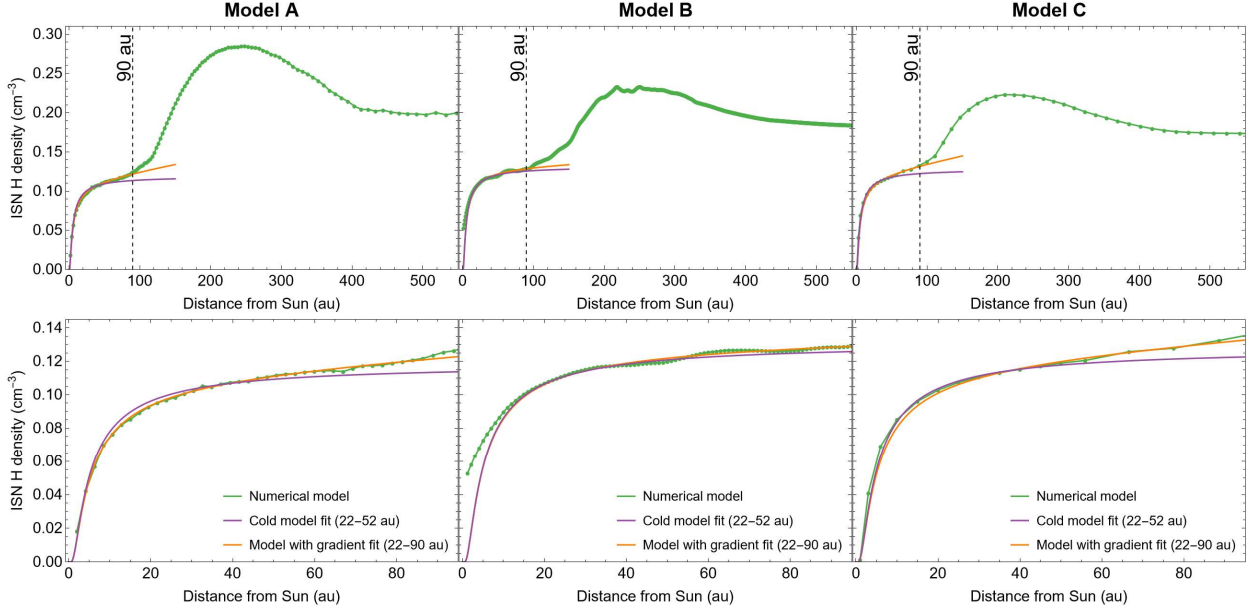

**Figure 2.** ISN hydrogen density along New Horizons' trajectory for the three models (in columns). The numerical results are presented using the green line. The cold model and the model with a gradient fitting the models over the limited range are shown using purple and orange lines, respectively. The bottom panel magnifies the portion within 90 au from the Sun for easier comparison.

The slope parameters found are positive, unlike the fit to the New Horizons' observations. These fits provide a good description of the density profile inside the termination shock beyond the ionization cavity. However, the discrepancies between the numerical models and the cold model extrapolations near 90 au show that the extrapolation of the ISN hydrogen density to the termination shock using the cold model to "tune" the global models is incorrect. Such an approach was commonly used (e.g., Zirnstein et al. 2016; Bzowski & Heerikhuisen 2020).

Using the ISN hydrogen density profiles predicted by the global models, we find scaling factors for each model that best fit the observations using $\chi^2$ minimization with the observational data. The obtained values are 1.0184, 0.9277, and 0.9484 for models A, B, and C, respectively. Figure 3 presents the scaled density profiles and their comparison with the observational data and the fits of the cold model and the model with gradient found in Section 2. Note that the model with gradient fit to the data has an opposite gradient, and thus, the extrapolation beyond the current range of New Horizons data differs significantly from the fits shown in Figure 2. Applying these factors to scale the VLISM ISN hydrogen density assumed as the boundary conditions, the corrected densities are 0.202, 0.168, and 0.163 cm$^{-3}$ for models A, B, and C, respectively. The mean value is ~0.18 cm$^{-3}$, i.e., about 10% less than the estimate of 0.195 cm$^{-3}$ from Swaczyna et al. (2020), which was based on the mean filtration factor provided by Müller et al. (2008). Considering non-linear effects, e.g., through coupling with the plasma, the best-fit densities may differ slightly from these estimates.

The density in the VLISM cannot be constrained directly. Therefore, we recommend a fiducial point along the New Horizons' trajectory at 40 au from the Sun, where the ISN hydrogen density is 0.11 cm$^{-3}$. The ecliptic coordinates are (285.62°, 1.94°). At this distance, the fits to the observed densities of all models considered in this study, including the analytic formulae, converge within less than 1%. This distance is far enough from the Sun that the consequences of the processes within the ionization cavity are negligible.

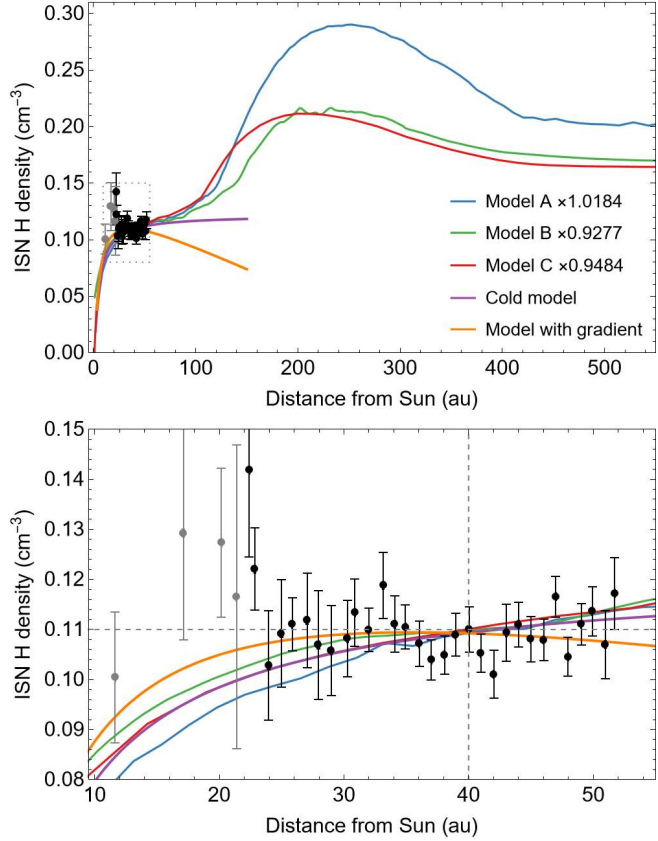

**Figure 3**. ISN hydrogen density profiles (color lines) from the fit analytic and scaled numerical models used in this study compared to the observations (dots with error bars). The bottom panel shows the density profiles within the range covered by the New Horizon observations. The fiducial point distance and density are marked with dashed lines (see text).

## 4. Discussion

Our analysis shows that the cold model reasonably approximates the ISN hydrogen density profile along the New Horizons' trajectory. However, the model with gradient, given by Equation (3), informed by the density profiles in the global heliosphere models, provides a statistically better fit. Interestingly, the best-fit slope of this additional gradient found from the fit to the New Horizons data has the opposite sign and is two times higher than the slopes from the global models.

To check if the discrepancy in the slope is connected with the temporal evolution of the ionization and radiation pressure, we compare the results of two time-dependent hot models. The first one is the Warsaw Test Particle Model (WTPM, Tarnopolski & Bzowski 2009; Sokół et al. 2015) implementing state-of-the-art models of the temporal evolution of the ionization processes (Sokół et al. 2020; Porowski et al. 2022) and radiation pressure (Kowalska-Leszczynska et al. 2018a, 2018b, 2020) along the trajectories of modeled atoms. The other is the LATMOS hot model (Lallement et al. 2005; Quémerais et al. 2006), which considers stationary ionization conditions applied with a year-to-year variation of the total ionization rate and radiation pressure based on the SWAN latitude-dependent data (Koutroumpa et al. 2019). Figure 4 compares these models with the data. We proportionally scaled both hot models to match the densities at the termination shock with the values obtained in our analysis. The figure shows a perfect agreement of the WTPM with the cold model. The LATMOS model predicts stronger variation as the annual ionization values are applied simultaneously along the atoms' trajectories. Therefore, we conclude that the simplifications of the cold model formula and the temporal variations of the ionization rates and radiation pressure are not the source of the discrepancy for the distance range covered in this study.

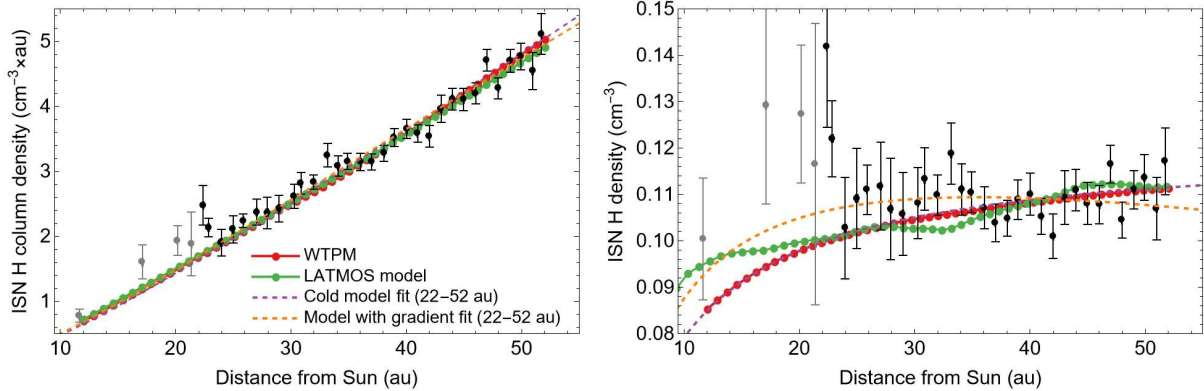

**Figure 4.** The ISN hydrogen density profiles predicted by the hot models and analytic formulas. The left and right panels show the column and local ISN hydrogen densities, respectively, as a function of distance from the Sun along the New Horizons' trajectory. The hot models are scaled by a constant factor to match the data. The hot models do not explain the negative slope.

The global models predict positive slopes with the magnitude depending on the detail of the modeling and the local models cannot explain the negative slope with time-depending ionization processes near the Sun. Consequently, this phenomenon requires a different explanation. The ISN hydrogen density may be related to temporal changes in the filtration at the outer heliospheric boundaries in response to the solar cycle and secular modulation of the solar wind (McComas et al. 2019). These changes influence the position of the heliopause by a few au throughout the solar cycle (e.g., Washimi et al. 2017; Izmodenov & Alexashov 2020; Sokół et al. 2021) and, thus, the filtration of ISN hydrogen in the heliosphere's outer boundaries. Another possibility is that the enhancement in the density is related to an inhomogeneity of the ISN hydrogen density in the VLISM.

## 5. Summary

New Horizons is the first spacecraft to provide PUI observations from the distant outer heliosphere, far beyond the ionization cavity. Consequently, the derived ISN hydrogen density profile is less affected by complex processes impacting the ISN hydrogen distribution within a few au from the Sun, such as ionization and radiation pressure. We find that the extended dataset of New Horizons observations follows the predictions of the cold model with the density far from the Sun $n_C = 0.125$ cm$^{-3}$. However, the model with an additional density gradient along the inflow direction provides a statistically better fit. Nevertheless, we found that the gradient coefficient is negative, i.e., the density decreases as the spacecraft moves toward the VLISM inflow direction. This finding is inconsistent with the global heliosphere models that predict an opposite sign of such gradient within the region of the supersonic solar wind.

We employ three global heliosphere models to put the New Horizons observations in the perspective of the ISN hydrogen density profile through the heliosphere's outer boundaries. We find scaling factors for each model that best fit the observations. The scaled profiles and the fits of the analytic models give very similar ISN hydrogen densities, 0.11 cm$^{-3}$ (±1%) at 40 au from the Sun, i.e., approximately midway through the used New Horizons' observations. We recommend this as a fiduciary point for future comparison with the predictions of the ISN hydrogen density models. The density at the termination shock is not a good fiduciary point because while the cold model prediction is close to the asymptotic value, the global models show strong gradients at these distances. The scaled ISN hydrogen density profiles from the global models using different boundary conditions estimate the ISN hydrogen density in the pristine VLISM between 0.163 and 0.202 cm$^{-3}$.

The apparent discrepancy between the global model predictions and the slope observed on New Horizons may result from transient effects, the filtration in the outer boundary regions of the heliosphere or a result of large-scale gradients in the pristine VLISM. Future observations of the SWAP instrument on New Horizons should allow further investigation of this problem.

## Acknowledgments

This research was supported by the International Space Science Institute (ISSI) in Bern through ISSI International Team project #541 (*Distribution of Interstellar Neutral Hydrogen in the Sun's Neighborhood*). P.S. is supported by a project co-financed by the Polish National Agency for Academic Exchange within the Polish Returns Programme (BPN/PPO/2022/1/00017) with a research component funded by the National Science Centre, Poland (2023/02/1/ST9/00004). M.B. and M.A.K. are supported by the National Science Centre, Poland (2019/35/B/ST9/01241). K.D. acknowledges support at JHU/APL by NASA under contracts NAS597271, NNX07AAJ69G, and NNN06AA01C and by subcontract at the Center for Space Research and Technology (Academy of Athens) and useful discussions within the SHIELD DRIVE Science Center: https://shielddrivecenter.com/. L.D. and H.R.M. are supported by NASA grant 80NSSC21K1681. F.F. is supported by NASA grants 80NSSC24K0267 and 80NSSC22K0524. F.F. also acknowledges the Texas Advanced Computing Center (TACC) at The University of Texas at Austin for providing HPC resources on Frontera supported by NSF award CISE-OAC-2031611. M.Z.K. and M.O. are supported by NASA grant 18-DRIVE18_0029, Our Heliospheric Shield, 80NSSC22M0164, and also by NASA HGI grant 80NSSC22K0525. D.K. is supported by CNES through its Sun Heliosphere Magnetosphere (SHM) program. I.K.L. is co-funded from the statutory funds of CBK PAN (own research fund TT-420). F.R. is funded by the IBEX mission as part of NASA's Explorer Program (80NSSC18K0237). The authors, except the corresponding author, are listed in alphabetical order. For the purpose of Open Access, the authors have applied a CC-BY public copyright license to any Author Accepted Manuscript (AAM) version arising from this submission.

## Data availability

The SWAP derived data products used in this study were obtained by McComas et al. (2021, 2022) and can be downloaded from: https://spacephysics.princeton.edu/missions-instruments/swap/pui-data-2021 and https://spacephysics.princeton.edu/missions-instruments/swap/pui-data-2022.